\documentclass[11pt]{article}

\usepackage{latexsym}
\usepackage{amsmath}
\usepackage{hyperref}
\usepackage{amssymb}
\usepackage{amsthm}
\usepackage[all]{xy}
\usepackage{color}
\usepackage{lscape}
\usepackage{graphicx}
\usepackage[margin=1in]{geometry}
\usepackage{algorithm}
\usepackage{algorithmic}

\def\Dmin{\lambda}

\def\cRatio{\gamma}
\def\cost{\textrm{cost}}
\def\prev{\textrm{prev}}

\title{Towards a General Framework for Searching on a Line\\ and Searching on $m$ Rays\footnote{This work was supported by FQRNT and NSERC.}}

\author{Prosenjit Bose,\\
Carleton University,\\
Ottawa, Canada\\
\texttt{jit@scs.carleton.ca}
\and
Jean-Lou De Carufel,\\
Carleton University,\\
Ottawa, Canada\\
\texttt{jdecaruf@cg.scs.carleton.ca}}

\date{}

\begin{document}
\newtheorem{definition}{Definition}
\newtheorem{theorem}{Theorem}
\newtheorem{proposition}{Proposition}
\newtheorem{lemma}{Lemma}
\newtheorem{corollary}{Corollary}
\newtheorem{conjecture}{Conjecture}

\maketitle

\begin{abstract}
Consider the following classical search problem:
given a target point $p\in \Re$,
starting at the origin,
find $p$ with minimum cost,
where cost is defined as the distance travelled.
Let $D=|p|$ be the distance of the point $p$ from the origin.
When no lower bound on $D$ is given, no competitive search strategy exists.
Demaine, Fekete and Gal
(\emph{Online searching with turn cost,
Theor. Comput. Sci., 361(2-3):342-355, 2006}) considered the situation where no lower bound on $D$ is given but a fixed \emph{turn cost} $t>0$ is charged every time the searcher changes direction. When the total cost is expressed as $\cRatio D + \phi$,
where $\cRatio$ and $\phi$ are positive constants, they showed that if $\cRatio$ is set to $9$, then the optimal search strategy has a cost of $9D+2t$. 
Although their strategy is optimal for $\cRatio = 9$, 
we prove that the minimum cost in their framework is $5 D + t + 2\sqrt{2D (2 D + t)} < 9D + 2t$. Note that the minimum cost requires knowledge of $D$.
However, given $D$, the optimal strategy has a smaller cost of $3D+t$.
Therefore, this problem cannot be solved optimally and exactly
when no lower bound on $D$ is given.

To resolve this issue,
we introduce a general framework where the cost of moving distance $x$ away from the origin is $\alpha_1 x+\beta_1$ and the cost of moving distance $y$ towards the origin is $\alpha_2 y + \beta_2$ for constants $\alpha_1, \alpha_2, \beta_1, \beta_2$. Given a lower bound $\lambda$ on $D$, we provide a provably optimal competitive search strategy when $\alpha_1, \alpha_2 ,\beta_1, \beta_2 \geq 0$ and $\alpha_1 + \alpha_2 > 0$. We show how our framework encompasses many of the results in the literature, and also point out its relation to other frameworks that have been proposed. 

Finally, we address the problem of searching for a target lying on one of $m$ rays extending from the origin where the cost is measured as the total distance travelled plus $t \geq 0$ times the number of turns. We provide a search strategy and compute its cost.
We prove our strategy is optimal for small values of $t$ and conjecture it is always optimal.
\end{abstract}

\newpage
\section{Introduction}
\label{section introduction}

Consider the following classical search problem: given a target point $p\in \Re$, starting at the origin, find $p$ with minimum cost, where cost is defined as the distance travelled. This problem and many of its variants have been extensively studied both in mathematics and computer science. For an encyclopaedic overview of the field, the reader is referred to the following books on the area~\cite{alpern2013book,alpern2003theory,gal1980}. Techniques developed to solve this family of problems have many applications in various fields such as robotics, scheduling, clustering, or routing to name a few~\cite{bose2004online,bose2001routing,dudek2010computational,lavalle2006planning,o2008comparing,pruhs2004online,zilberstein2003optimal}. In particular, solutions to these problems have formed the backbone of many competitive online algorithms (see~\cite{DBLP:journals/sigact/ChrobakK06} for a comprehensive overview). 
 
Demaine, Fekete and Gal~\cite{DBLP:journals/tcs/DemaineFG06}
studied a variant of the classical problem where in addition to the distance travelled, a cost of $t>0$ is charged for each turn.
Their total search cost is expressed as $\cRatio D + \phi$,
where $\cRatio$ and $\phi$ are constants, and $D=|p|$ is the distance of the point $p$ from the origin\footnote{There does not seem to be a consensus on notation for the linear search problem.
For instance,
in~\cite{DBLP:journals/tcs/Lopez-OrtizS01},
$D$ represents an upper bound on the distance.
In~\cite{DBLP:journals/tcs/DemaineFG06},
$d$ represents the turn cost.
As for the distance of $p$ from the origin
it is denoted by $|H|$ in~\cite{alpern2013book,alpern2003theory},
by $n$ in~\cite{DBLP:journals/iandc/Baeza-YatesCR93}
and it does not have any specific notation in~\cite{DBLP:journals/tcs/Lopez-OrtizS01}.}. They present a competitive search algorithm despite their assumption that no lower bound on $D$ is given. Normally, with no lower bound given on $D$, a search algorithm cannot be competitive. If an adversary places $p$ at a distance $\epsilon>0$ from the origin and the first step taken by the algorithm is $\delta>0$ in the wrong direction, the ratio $\delta/\epsilon$ cannot be bounded. They circumvent this issue by leveraging the fact that an optimal search strategy will have to turn at least once in the worst case.
They prove that if $\cRatio = 9$,
the optimal strategy is the following. Let $x_i = \frac{1}{2}\left(2^i-1\right)t$ ($i \geq 1$). At step $i$, if $i$ is even, move to $x_i$ and then return to the origin. If $i$ is odd, move to $-x_i$ and then return to the origin. The total cost of this strategy is $9D + 2t$ in the worst case.
Notice that their strategy is defined only for $t > 0$ since when $t=0$, their search algorithm remains at the origin since $x_i =0$. Moreover, they only studied the case where $\cRatio = 9$.
There is no guarantee that setting $\cRatio = 9$ yields an optimal solution.
There could be a solution with $\cRatio D + \phi < 9D + 2t$,
where $\cRatio \neq 9$.
In their discussion on the trade-off between $\cRatio$ and $\phi$,
they write~\cite[p.351]{DBLP:journals/tcs/DemaineFG06}:
``We have not yet characterized it analytically, though we expect that to be possible.''

This is what initiated our investigation. 
In Section~\ref{section previous work},
we establish an analytic characterization of the trade-off between $\cRatio$ and $\phi$.
We do so by further refining the techniques from~\cite{DBLP:journals/tcs/DemaineFG06},
which allows us to prove that when $\cRatio = 9$,
the solution is not optimal.
Given our analytic characterization, we prove that the minimum cost is $5 D + t + 2\sqrt{2D (2 D + t)} < 9D + 2t$. The caveat is that the minimum cost requires knowledge of $D$. However, given $D$, the optimal strategy has a smaller cost of $3D+t$.
Therefore, this problem cannot be solved optimally when no lower bound on $D$ is known. 

In order to fill some of these gaps in the model, we first study the situation where one is given a lower bound of $\Dmin$ on $D$, and a search algorithm is charged the total distance travelled as well as $t>0$ for each turn. Surprisingly, in this setting, we prove that when $t/2\Dmin<1$, the optimal search cost is still $9D$ which is the optimal search cost when $t=0$ as shown by Baeza-Yates et al.~\cite{DBLP:journals/iandc/Baeza-YatesCR93}. When $t/2\Dmin\geq 1$, the optimal search cost is $2\frac{\left(\frac{t}{2\Dmin}+2\right)\left(\frac{t}{2\Dmin}+\frac{1}{2}\right)}{\frac{t}{2\Dmin}} \, D$.
Alpern and Gal~\cite[Section 8.4]{alpern2003theory} considered this framework for solving the problem of searching on a line with turn cost.
They provided a strategy with expected cost $9D + 2t$.
Then,
the question of whether this is optimal was left as open.
Our result is an optimal deterministic solution to this problem.
In Section~\ref{section different work},
we develop our own tools,
from which we analyze and characterize an optimal deterministic solution.

Moreover, we introduce a more general framework
(see Section~\ref{section general framework searching line})
where the cost of moving distance $x$ away from the origin is $\alpha_1 x+\beta_1$ and the cost of moving distance $y$ towards the origin is $\alpha_2 y + \beta_2$ for constants $\alpha_1, \alpha_2, \beta_1, \beta_2$. Given a lower bound $\Dmin$ on $D$, we provide a provably optimal competitive search strategy when $\alpha_1, \alpha_2 , \beta_1, \beta_2 \geq 0$ and $\alpha_1 + \alpha_2 > 0$.
We show how our framework encompasses many of the results in the literature, and also point out its relation to other frameworks that have been proposed such as the one proposed by Gal~\cite{gal72,gal1980}.

Finally, in Section~\ref{section searching m rays turn cost},
we address the problem of searching for a target lying on one of $m$ rays extending from the origin where the cost is measured as the total distance travelled plus $t \geq 0$ times the number of turns. We provide a search strategy and compute its cost.
We prove our strategy is optimal for small values of $t$ and conjecture it is always optimal. In fact, surprisingly, the ratio $t/2\Dmin$ plays a significant role. When this ratio is small, we show that the cost of our search strategy is identical to the cost of the optimal strategy with no turn cost. This is why we can claim optimality for small values of $t$.

\section{Previous Work}
\label{section previous work}

A \emph{search strategy} for the problem of searching on a line is a function
$\mathcal{S}(i) = (x_i,r_i)$ defined for all integers $i \geq 1$.
At step $i$,
the searcher travels a distance of $x_i$
on ray $r_i \in \{\textrm{left},\textrm{right}\}$.
If he does not find the target,
he goes back to the origin
and proceeds with step $i+1$.
Let $D$ be the distance between the searcher and the target at the beginning of the search.
Traditionally,
the goal is to find a strategy $\mathcal{S}$
that minimizes the \emph{competitive ratio} $CR(\mathcal{S})$ (or \emph{competitive cost}) defined as the total distance travelled by the searcher divided by $D$,
in the worst case.
If $D$ is given to the searcher,
any strategy $\mathcal{S}$ such that $\mathcal{S}(1) = (D,\textrm{left})$ and $\mathcal{S}(2) =(D,\textrm{right})$ is optimal with a competitive ratio of $3$ in the worst case.
If $D$ is unknown,
a lower bound $\Dmin \leq D$ must be given to the searcher,
otherwise the competitive ratio is unbounded in the worst case.

Let $Left = \{i\mid r_i = \textrm{left}\}$
and $Right = \{i\mid r_i = \textrm{right}\}$.
To guarantee that,
wherever the target is located,
we can find it with a strategy $\mathcal{S}$,
we must have $\sup_{i\in Left} x_i = \sup_{i\in Right} x_i = \infty$.
We say that $\mathcal{S}$ is \emph{monotonic}
if the sequences $\left(x_i\right)_{i\in Left}$
and $\left(x_i\right)_{i\in Right}$
are strictly increasing.
The strategy $\mathcal{S}$ is said to be \emph{periodic}
if $r_1\neq r_2$ and $r_i = r_{i+2}$ for all $i\geq 1$.
We know from previous work
(see~\cite{gal1980,DBLP:journals/iandc/Baeza-YatesCR93} for instance)
that there is an optimal strategy that is periodic and monotonic.
Let us say that a strategy is \emph{fully monotonic}
if the sequence $\left(x_i\right)_{i\geq 1}$ is monotonic and non-decreasing.
We prove the following theorem in appendix~\ref{section monotonic periodic}
\begin{theorem}
\label{thm general periodic monotonic m = 2}
Let $\cost_1(x) = \alpha_1 x+\beta_1$ be the cost of walking distance $x$ away from the origin
and $\cost_2(y) = \alpha_2 y+\beta_2$ be the cost of walking distance $y$ towards the origin.
If $\alpha_1 \geq 0$, $\alpha_2\geq 0$, $\alpha_1 + \alpha_2 > 0$, $\beta_1 \geq 0$ and $\beta_2 \geq 0$,
there exists an optimal search strategy that is periodic and fully monotonic.
\end{theorem}

For the problem of searching on a line,
we have $\alpha_1 = \alpha_2 = 1$ and $\beta_1 = \beta_2 = 0$.
If a fixed cost of $t$ is charged to the searcher every time he changes direction,
we have $\alpha_1 = \alpha_2 = 1$, $\beta_1 = 0$ and $\beta_2 = t$.
We call this search problem \emph{Searching on a line with Turn Cost}.
If we find the smallest $\cRatio$ for which there exists
a periodic and fully monotonic strategy with
%\begin{align*}
%\inf_{\varepsilon > 0}\frac{(\alpha_1 x_1+\beta_1)+(\alpha_2 x_1+\beta_2)+(\alpha_1(\Dmin + %\varepsilon)+\beta_1)}{\Dmin + \varepsilon} &\leq \cRatio, \\
%\sup_{k\geq 1}\inf_{\varepsilon > 0}\frac{\sum_{i = 1}^{k+1}\Big((\alpha_1 %x_i+\beta_1)+(\alpha_2 x_i+\beta_2)\Big)+(\alpha_1(x_k+\varepsilon)+\beta_1)}{x_k+\varepsilon} %&\leq \cRatio
%\end{align*}
%for any $\varepsilon > 0$,
%which is equivalent to
\begin{align*}
\frac{(\alpha_1 + \alpha_2)x_1 + \beta_1 + \beta_2+\alpha_1\Dmin+\beta_1}{\Dmin} \leq \cRatio \;\;\textrm{and}\;\;
\sup_{k\geq 1}\frac{\sum_{i = 1}^{k+1}\Big((\alpha_1 +\alpha_2)x_i+\beta_1+\beta_2\Big)+\alpha_1 x_k+\beta_1}{x_k} \leq \cRatio,
\end{align*}
then we are done by Theorem~\ref{thm general periodic monotonic m = 2}.

\subsection{The General Framework of Gal}
\label{subsection framework gal}

Gal~\cite{gal72,gal1980} proved a general inequality
that helps computing lower bounds on the competitive cost of optimal strategies ---for a general class of Searching-on-$m$-Rays problems--- in the worst case
(refer to Theorem 1A of~\cite[Section 4]{gal72} or to Corollary 1 of~\cite[Chapter 6]{gal1980}).
\begin{theorem}[Gal, 1972, 1980]
\label{thm gal}
Let $(x_i)_{i\in\mathbb{Z}}$ be a sequence of positive real numbers.
Let $k_0$ and $L$ be positive integers.
Let $(\beta_j)_{-L \leq j \leq L}$ be a sequence of non-negative real numbers
and $(\alpha_j)_{-\infty < j < \infty}$ be a sequence of real numbers such that $\alpha_j \geq 0$ for all $|j| > k_0$. Then
\begin{align}
\label{thm ineq gal}
\limsup_{\stackrel{i \rightarrow \infty}{i \in \mathbb{Z}}} \frac{\sum_{j=-\infty}^{\infty}\alpha_j x_{i+j}}{\sum_{j=-L}^{L}\beta_j x_{i+j}} \geq \inf_{a \geq 0} \frac{\sum_{j=-\infty}^{\infty}\alpha_j a^j}{\sum_{j=-L}^{L}\beta_j a^j} .
\end{align}
\end{theorem}

Gal solved the problem of searching on $m$ rays for any $m \geq 2$
(refer to Chapters 7 and 8 of~\cite{gal1980}) using Theorem~\ref{thm gal}.
For instance,
here is how we solve the classical problem of searching on a line.
Let $k_0 = L = 1$, $\beta_{-1} = \beta_1 = 0$, $\beta_0 = 1$, 
$\alpha_j = 1$ ($j \leq 1$) and $\alpha_j = 0$ ($j > 1$).
From~\eqref{thm ineq gal},
we find
$$\sup_{k\in\mathbb{Z}}\frac{\sum_{j=-\infty}^{k+1}2x_j+x_k}{x_k} = 1+2\sup_{k\in\mathbb{Z}}\frac{\sum_{j=-\infty}^1x_{j+k}}{x_{0+k}} \geq 1+2\inf_{a \geq 0} \frac{\sum_{j=-\infty}^1 a^j}{a^0} = 9 .$$
Since $x_i = 2^i \Dmin$
has a competitive cost of $9$ in the worst case, we are done
(refer to Gal~\cite{gal1980} for further details about how to handle the lower bound on $D$).

We cannot use Theorem~\ref{thm gal} for the problem of searching on a line with turn cost.
At each step,
we need to add $t$ to the total cost.
Therefore,
in~\eqref{thm ineq gal},
``$\alpha_j x_{i+j}$''
should be replaced by ``$\alpha_j x_{i+j} + t$''.
Unfortunately,
since the summation is infinite,
this would cause the series to diverge.
In that sense,
the family of Searching-on-a-Line problems we are considering in Theorem~\ref{thm general periodic monotonic m = 2}
is more general.
In Section~\ref{section general framework searching line},
we explain how to solve optimally any
Searching-on-a-Line problems that are considered
in Theorem~\ref{thm general periodic monotonic m = 2}.

\subsection{Previous Work with Turn Cost}
\label{subsection previous work turn cost}

The problem of searching on a line with turn cost was studied in a different framework by Demaine et al.~\cite{DBLP:journals/tcs/DemaineFG06}
than the one introduced at the beginning of Section~\ref{section previous work}.
We first note that,
given a turn cost of $t$,
if we know $D$,
the optimal strategy is still $x_i = D$
and it has a competitive cost of
$\frac{3D+t}{D} = 3 + \frac{t}{D} \leq 3+\frac{t}{\Dmin}$
in the worst case.
The result stands for any value of $t$,
including when $t = 0$,
where we get a competitive cost of $3 + \frac{0}{D} = 3$ in the worst case.
This corresponds to the case where $D$ is known and there is no turn cost.

Demaine et al.~\cite{DBLP:journals/tcs/DemaineFG06}
expressed the total cost of a search as $\cRatio D + \phi$,
where $\cRatio$ and $\phi$ are constants,
and they suppose that no lower bound is given to the searcher.
Their goal is to minimize the total cost of the search in the worst case.
They proved that if $\cRatio = 9$,
the optimal strategy is $x_i = \frac{1}{2}\left(2^i-1\right)t$ ($i \geq 1$)
with a total cost of $9D + 2t$ in the worst case.
This strategy is defined for any $t > 0$.
For $t = 0$,
we get $x_i = 0$
($i \geq 1$),
which is not a search strategy since the searcher does not move.
This is unfortunate since
we would like to have a strategy that depends on $t$ and that is valid when we set $t = 0$.

Since they studied the case where $\cRatio = 9$,
there is no guarantee that $9D + 2t$ is optimal.
There could be a solution with $\cRatio D + \phi < 9D + 2t$,
where $\cRatio \neq 9$.
They mention that the trade-off between $\cRatio$ and $\phi$ should be looked at more closely.
In this section,
we establish an analytic characterization of the trade-off between $\cRatio$ and $\phi$.
The analysis of that characterization raises significant questions about the model of Demaine et al.
(refer to Section~\ref{subsubsection tradeoff}).

Moreover,
the fact that Baeza-Yates et al.~\cite{DBLP:journals/iandc/Baeza-YatesCR93} and Demaine et al. work in different frameworks makes the strategies difficult to compare.
If we suppose that no lower bound is given in the model of Baeza-Yates et al, then the competitive cost is unbounded in the worst case.
If we suppose that a lower bound is given in the model of Demaine et al.,
then the optimal strategy might be different since an extra information is given to the searcher.
Previous results
(see~\cite{DBLP:conf/esa/BoseCD13,DBLP:journals/tcs/Lopez-OrtizS01} for instance)
that generalize the work of Baeza-Yates et al. aim at optimizing the same cost function,
namely the competitive cost,
and the comparison with the results of Baeza-Yates et al. is immediate.

\subsubsection{The Trade-off Between $\cRatio$ and $\phi$}
\label{subsubsection tradeoff}

Let us write $\phi$ as a function $\phi(\cRatio)$ that depends on $\cRatio$.
For each value of $\cRatio$,
there is an optimal value for $\phi$ that  minimizes $\cRatio D + \phi(\cRatio)$.
Demaine et al.~\cite{DBLP:journals/tcs/DemaineFG06} showed that $\phi(9) = 2t$.
We want to find the value of $\phi(\cRatio)$ for any $\cRatio > 9$.
Using the technique developed by Demaine et al.,
we prove the following theorem.
\begin{theorem}
\label{thm demaine complete}
The optimal strategy given any $\cRatio \geq 9$ is
$$x_i = \frac{1}{2} \left(-1 + \left(\frac{\cRatio-1 - \sqrt{(\cRatio - 1) (\cRatio - 9)}}{4}\right)^i\right)t ,$$
with total cost
$\cRatio D + \frac{1}{4} \left(\cRatio - 1 - \sqrt{(\cRatio-1)(\cRatio-9)} \right) t $
in the worst case.
\end{theorem}
In other words,
\begin{align}
\label{def phi}
\phi(\cRatio) = \frac{1}{4} \left(\cRatio - 1 - \sqrt{(\cRatio-1)(\cRatio-9)} \right) \, t .
\end{align}

\proof
Following~\cite{DBLP:journals/tcs/DemaineFG06},
the infinite linear program we need to solve is
$\min \phi(\cRatio)$
subject to
$$
\renewcommand{\tabcolsep}{1.75pt}
\left\{
\begin{tabular}{rrllllll}
$2x_1$           $\phantom{+}$&               &     & & & & $+$ $\phantom{1}t$ & $\leq$ $\phi(\cRatio)$ \\
$(3-\cRatio)x_1$ $+$ & $2x_2$           &     & & & & $+$ $2t$ & $\leq$ $\phi(\cRatio)$ \\
$2x_1$           $+$ & $(3-\cRatio)x_2$ &$+$ $2x_3$ &           &                     &       & $+$ $3t$ & $\leq$ $\phi(\cRatio)$ \\
               &               &     &           &                     &       & & $\vdots$ \\
$2x_1$           $+$ & $2x_2$           &$+$ $\cdots$  &$+$ $2x_{n-2}$ &$+$ $(3-\cRatio)x_{n-1}$ &$+$ $2x_n$ & $+$ $nt$ & $\leq$ $\phi(\cRatio)$ \\
               &               &     &           &                     &       & & $\vdots$ \\
\end{tabular}
\right.
$$
and $x_i \geq 0$ for all $i \geq 1$.

We use the following dual multipliers\footnote{The
optimal solutions to all known variants of the line searching problem
have to do with exponential strategies.
By looking for dual multipliers of exponential forms,
we obtained an optimal solution.}:
$y_i = r_1^{-i}(\cRatio)$ ($i\geq 1$),
where
\begin{align}
\label{eqn def r1 r2}
r_1(\cRatio) = \frac{\cRatio - 1 + \sqrt{(\cRatio - 9)(\cRatio - 1)}}{4} , \qquad
r_2(\cRatio) = \frac{\cRatio - 1 - \sqrt{(\cRatio - 9)(\cRatio - 1)}}{4}
\end{align}
are the two roots of $x^2 - \frac{\cRatio - 1}{2} x + \frac{\cRatio - 1}{2} = 0$.
Therefore,
the dual system successively becomes
\begin{align*}
\left(\sum_{i = 1}^{\infty}2y_i - (\cRatio - 1)y_2\right)x_1
+ \cdots + \left(\sum_{i = \ell}^{\infty}2y_i - (\cRatio - 1)y_{\ell+1}\right)x_{\ell} + \cdots + \left(\sum_{i = 1}^{\infty}iy_i \right)t &\leq  \left(\sum_{i = 1}^{\infty}y_i \right)\phi(\cRatio) ,\\
0 + 0 + \cdots + 0 + \cdots + \frac{r_1(\cRatio)}{(r_1(\cRatio) - 1)^2}t &\leq \frac{1}{r_1(\cRatio) - 1}\phi(\cRatio) ,\\
r_2(\cRatio)t &\leq \phi(\cRatio) .
\end{align*}
We prove that $r_2(\cRatio)t$ is optimal by considering
$x_i = \frac{1}{2}\left(r_2^i(\cRatio) - 1\right)t$.
Then,
$$\sum_{i=1}^{\ell}2x_i - (\cRatio - 1)x_{\ell-1} + \ell t \leq \phi(\cRatio)$$
becomes
$r_2(\cRatio) t \leq \phi(\cRatio)$.
\qed

The trade-off between $\cRatio$ and $\frac{1}{t}\phi(\cRatio)$
is depicted in Figure~\ref{figure tradeoff} of Appendix~\ref{section figures}.
That figure corresponds to the one presented by Demaine et al.~\cite[Figure 2]{DBLP:journals/tcs/DemaineFG06}.
Moreover,
Demaine et al. conjecture that $\lim_{\cRatio\rightarrow \infty}\frac{1}{t}\phi(\cRatio) = 1$.
From~\eqref{def phi},
that conjecture becomes a straightforward calculus exercise.

We turn to the question of minimizing the total cost of $x_i$
in the worst case,
where the total cost $TC(x_i) = \cRatio D + r_2(\cRatio) t $
is a function of $\cRatio$.
We first differentiate $TC(x_i)$ with respect to $\cRatio$
and we solve
$\frac{d}{d\cRatio}TC(x_i) = D - \frac{r_2(\cRatio) - 1}{\cRatio-1 - 4r_2(\cRatio)}t = 0$.
We find
$\cRatio = 5 + 2\frac{4D+t}{\sqrt{2D(2D+t)}}$.
Moreover,
from elementary calculus,
we get
$\frac{d^2}{d\cRatio^2}TC(x_i) = 2\frac{(\cRatio - 3 - 2 r_2(\cRatio)) (r_2(\cRatio) - 1) }{(\cRatio - 1 - 4 r_2(\cRatio))^3} t \geq 0 $
for all $\cRatio \geq 9$.
Therefore,
the optimal total cost is
$$\left(5 + 2\frac{4D+t}{\sqrt{2D(2D+t)}}\right)D + r_2\left(5 + 2\frac{4D+t}{\sqrt{2D(2D+t)}}\right)t = 5 D + t + 2\sqrt{2D (2 D + t)}$$
in the worst case.
Moreover,
$$x_i=\frac{1}{2} \left(\left(1 + 2\frac{D}{\sqrt{2D (2 D + t)}}\right)^i - 1\right) t$$
is the corresponding optimal strategy.
Notice that
$9D+2t \geq 5 D + t + 2\sqrt{2D (2 D + t)}$
for all $D \geq 0$,
with equality if and only if $t = 0$.
However,
the strategy
$$x_i = \frac{1}{2} \left(-1 + \left(\frac{\cRatio-1 - \sqrt{(\cRatio - 1) (\cRatio - 9)}}{4}\right)^i\right)t ,$$
is not defined for $t = 0.$
This shows that $9D+2t$ is not the optimal total cost in the worst case.
Furthermore,
the strategy $x_i$ that minimizes $CR(x_i)$ depends on $D$.
But if we know $D$, there is a better strategy,
which we described at the beginning of the section.

\section{A different Framework}
\label{section different work}

In view of the discussion of the previous section,
we want to address the problem in a different framework.
We suppose the following: (a) We want to minimize the competitive cost $\cRatio$ of the search in the worst case
(as in the classical problem of searching on a line), (b) a lower bound $\Dmin \leq D$ is given to the searcher, (c) each time we change direction, we need to pay $t$, (d) the optimal strategy depends only on $t$ and $\Dmin$, (e) the optimal strategy is defined for any $t \geq 0$ and (f) when $t = 0$,
the optimal strategy has a competitive cost of $9$ in the worst case.

This framework generalizes the one of Baeza-Yates~\cite{DBLP:journals/iandc/Baeza-YatesCR93}
and it encapsulates what Demaine et al.~\cite{DBLP:journals/tcs/DemaineFG06} studied.
We first show that if $\frac{t}{2\Dmin} \leq 1$,
a competitive cost of $9$ is still achievable in the worst case.
\begin{lemma}
\label{lemma turn cost c = 9}
If $\frac{t}{2\Dmin} \leq 1$,
the strategy
$$x_i = \left(\left(\left(1 - \frac{t}{2\Dmin}\right)i + \left(1 + \frac{t}{2\Dmin} \right)\right) 2^i - \frac{t}{2\Dmin}\right)\Dmin $$
is optimal and has a competitive cost of $9$.
\end{lemma}

\proof
If $\frac{t}{2\Dmin} \leq 1$,
then
$\frac{(2x_1+t)+\Dmin}{\Dmin} = 9$
and for all $n\geq 2$,
$$\frac{\sum_{i=1}^n (2x_i + t) + x_{n-1}}{x_{n-1}} = 9 .$$
The strategy $x_i$ cannot do better than the optimal strategy for searching on a line.
Therefore,
$x_i$ is optimal.
\qed

When $t=0$,
we find $x_i = (i+1)2^i\Dmin$,
which has a competitive cost of $9$ in the worst case.
In Appendix~\ref{section proof lemma},
we prove the following lemma.
\begin{lemma}
\label{lemma > 9}
Let $\cRatio$ be the competitive cost of the optimal strategy ---for searching on a line with turn cost--- in the worst case.
If $\frac{t}{2\Dmin} > 1$,
then $\cRatio > 9$.
\end{lemma}

\begin{theorem}
\label{thm turn cost with m = 2}
If $\frac{t}{2\Dmin} \geq 1$,
the strategy
$$x_i = \left(\left(1+\frac{t}{2\lambda}\right)\left(1 + \left(\frac{t}{2\lambda}\right)^{-1}\right)^i  - \frac{t}{2\Dmin}\right)\Dmin .$$
is optimal and has a competitive cost of
$\frac{2x_1+t+\Dmin}{\Dmin} = 2\frac{\left(\frac{t}{2\Dmin}+2\right)\left(\frac{t}{2\Dmin}+\frac{1}{2}\right)}{\frac{t}{2\Dmin}}$.
\end{theorem}

The shape of the competitive cost of an optimal strategy in the worst case is depicted in Figure~\ref{figure optimal with turn cost} of Appendix~\ref{section figures}.

\proof
We have
$\frac{2x_1+t+\Dmin}{\Dmin} = 2\frac{\left(\frac{t}{2\Dmin}+2\right)\left(\frac{t}{2\Dmin}+\frac{1}{2}\right)}{\frac{t}{2\Dmin}}$
and for all $n\geq 2$,
$$\frac{\sum_{i=1}^n (2x_i + t) + x_{n-1}}{x_{n-1}} = 2\frac{\left(\frac{t}{2\Dmin}+2\right)\left(\frac{t}{2\Dmin}+\frac{1}{2}\right)}{\frac{t}{2\Dmin}} .$$
Moreover,
if $\frac{t}{2\Dmin} = 1$,
then
$2\frac{\left(\frac{t}{2\Dmin}+2\right)\left(\frac{t}{2\Dmin}+\frac{1}{2}\right)}{\frac{t}{2\Dmin}} = 9$.
For the rest of the proof,
we suppose that $\frac{t}{2\Dmin} > 1$.

Let $\cRatio$ be the competitive cost of an optimal strategy $(x_i)_{i=1}^{\infty}$.
By Lemma~\ref{lemma > 9},
$\cRatio > 9$.
Also,
we know that $x_i$ satisfies
\begin{align}
\label{basic ineq x1}
\frac{(2x_1 + t) + \Dmin}{\Dmin} \leq \cRatio
\end{align}
and
\begin{align}
\label{basic ineq xi}
\frac{\sum_{i=1}^{n+1} (2x_i + t) + x_n}{x_n} \leq \cRatio
\end{align}
for all $n\geq 1$.
Therefore,
\begin{align}
\label{ineq constraint x1}
0 < \Dmin \leq x_1 \leq \frac{(\cRatio - 1)\Dmin - t}{2}
\end{align}
and for all $n \geq 1$,
\begin{align}
\label{eqn general equiv}
x_{n+1} \leq A x_n - B - \sum\limits_{i=1}^{n-1} (x_i + C),
\end{align}
where $A = \frac{\cRatio - 3}{2}$, $B = t$ and $C = \frac{t}{2}$.
We can prove by induction that
for all $0 \leq m \leq n-1$,
\begin{align}
\label{eqn f n+1 m}
x_{n+1} \leq \tau_m x_{n-m} - \mu_m - \nu_m\sum\limits_{i=1}^{n-1 - m} (x_i + C),
\end{align}
where
\begin{align*}
\tau_0 &= A , & \mu_0 &= B , & \nu_0 &= 1 ,\\
\tau_{m+1} &= \tau_m A - \nu_m , & \mu_{m+1} &= \tau_m B + \mu_m + \nu_m C , & \nu_{m+1} &= \tau_m + \nu_m .
\end{align*}
From the theory of characteristic equations,
we find
\begin{align*}
\tau_m = & \frac{r_1(\cRatio)(r_1(\cRatio)-1)}{r_1(\cRatio)-r_2(\cRatio)} \, r_1^n(\cRatio) - \frac{r_2(\cRatio)(r_2(\cRatio)-1)}{r_1(\cRatio)-r_2(\cRatio)} \, r_2^n(\cRatio) ,\\
\mu_m = & \frac{r_1(\cRatio)(2r_1(\cRatio)-1)}{2(r_1(\cRatio)-1)(r_1(\cRatio)-r_2(\cRatio))} \, t\, r_1^n(\cRatio) - \frac{r_2(\cRatio)(2r_2(\cRatio)-1)}{2(r_2(\cRatio)-1)(r_1(\cRatio)-r_2(\cRatio))} \, t\, r_2^n(\cRatio) + \frac{t}{2} ,\\
\nu_m = & \frac{r_1(\cRatio)}{r_1(\cRatio)-r_2(\cRatio)} \, r_1^n(\cRatio) - \frac{r_2(\cRatio)}{r_1(\cRatio)-r_2(\cRatio)} \, r_2^n(\cRatio) ,
\end{align*}
where $r_1(\cdot)$ and $r_2(\cdot)$ are defined as in~\eqref{eqn def r1 r2}.

From~\eqref{eqn f n+1 m} with $m:=n-1$,
we have
\begin{align}
\label{ineq Klein}
x_{n+1} \leq \tau_{n-1} x_1 - \mu_{n-1}
\end{align}
for all $n \geq 1$.
By Theorem~\ref{thm general periodic monotonic m = 2},
$x_{n+1}$ is increasing with respect to $n$.
Also,
$x_{n+1}$ is unbounded.
Therefore,
$\delta(n) = \tau_{n-1} x_1 - \mu_{n-1}$
must be unbounded.

Let us study $\delta(n)$,
which can be written as
$\delta(n) = a(\cRatio, t) \, r_1^n(\cRatio) - b(\cRatio, t) \, r_2^n(\cRatio) + c(\cRatio,t)$,
where
\begin{align*}
a(\cRatio, t) &=  \frac{r_1(\cRatio)(r_1(\cRatio)-1)}{r_1(\cRatio)-r_2(\cRatio)} \,x_1
- \frac{r_1(\cRatio)(2r_1(\cRatio)-1)}{2(r_1(\cRatio)-1)(r_1(\cRatio)-r_2(\cRatio))} \, t,\\
b(\cRatio, t) &= \frac{r_2(\cRatio)(r_2(\cRatio)-1)}{r_1(\cRatio)-r_2(\cRatio)} \, x_1
- \frac{r_2(\cRatio)(2r_2(\cRatio)-1)}{2(r_2(\cRatio)-1)(r_1(\cRatio)-r_2(\cRatio))} \, t,\\
c(\cRatio, t) &=  - \frac{t}{2}.
\end{align*}

We prove that $b(\cRatio, t) < 0$ and $a(\cRatio, t) \geq 0$,
for all $\cRatio > 9$ and $t > 2\Dmin$.
We start with $b(\cRatio, t) < 0$.
We have
\begin{align}
\nonumber
& \, b(\cRatio, t) \\
\nonumber
=& \, \frac{r_2(\cRatio)(r_2(\cRatio)-1)}{r_1(\cRatio)-r_2(\cRatio)} \, x_1
- \frac{r_2(\cRatio)(2r_2(\cRatio)-1)}{2(r_2(\cRatio)-1)(r_1(\cRatio)-r_2(\cRatio))} \, t \\
\nonumber
\leq& \, \frac{r_2(\cRatio)(r_2(\cRatio)-1)}{r_1(\cRatio)-r_2(\cRatio)} \, \frac{(\cRatio - 1)\Dmin - t}{2}
- \frac{r_2(\cRatio)(2r_2(\cRatio)-1)}{2(r_2(\cRatio)-1)(r_1(\cRatio)-r_2(\cRatio))} \, t &&\textrm{since $1 < r_2(\cRatio) < r_1(\cRatio)$}\\
\nonumber
&&&\textrm{for all $\cRatio > 9$ and by~\eqref{ineq constraint x1},}\\
\nonumber
=& \, \frac{r_2(\cRatio)\left(((\cRatio-1)\Dmin-t) \, r_2^2(\cRatio) -2(\cRatio-1)\Dmin \, r_2(\cRatio) +(\cRatio - 1)\Dmin \right)}{2(r_2(\cRatio)-1)(r_1(\cRatio)-r_2(\cRatio))} \\
\label{last line b2 negative}
=& \, \frac{r_2(\cRatio)(\cRatio - 1)\left( ((\cRatio - 5)\Dmin - t) \, r_2(\cRatio) - ((\cRatio - 3)\Dmin - t) \right)}{4(r_2(\cRatio)-1)(r_1(\cRatio)-r_2(\cRatio))}
&&\textrm{by~\eqref{eqn def r1 r2}.}
\end{align}
We prove that~\eqref{last line b2 negative} is negative
by looking at two cases:
(1) $(\cRatio - 5)\Dmin - t \geq 0$
or (2) $(\cRatio - 5)\Dmin - t < 0$.
\begin{enumerate}
\item If $(\cRatio - 5)\Dmin - t \geq 0$,
then $(\cRatio - 3)\Dmin - t > (\cRatio - 5)\Dmin - t \geq 0$.
Therefore,
using elementary calculus,
we can prove
$$1 < r_2(\cRatio) < \frac{(\cRatio - 3) - 2}{(\cRatio - 5) - 2} = \frac{(\cRatio - 3)\Dmin - 2\Dmin}{(\cRatio - 5)\Dmin - 2\Dmin} < \frac{(\cRatio - 3)\Dmin - t}{(\cRatio - 5)\Dmin - t}$$
for all $\cRatio > 9$ and $t > 2\Dmin$.
Therefore,
\begin{align*}
& \, \frac{r_2(\cRatio)(\cRatio - 1)\left( ((\cRatio - 5)\Dmin - t) \, r_2(\cRatio) - ((\cRatio - 3)\Dmin - t) \right)}{4(r_2(\cRatio)-1)(r_1(\cRatio)-r_2(\cRatio))} \\
<& \, \frac{r_2(\cRatio)(\cRatio - 1)\left( ((\cRatio - 5)\Dmin - t) \, \frac{(\cRatio - 3)\Dmin - t}{(\cRatio - 5)\Dmin - t} - ((\cRatio - 3)\Dmin - t) \right)}{4(r_2(\cRatio)-1)(r_1(\cRatio)-r_2(\cRatio))} = 0 .
\end{align*}

\item If $(\cRatio - 5)\Dmin - t < 0$,
then we use the fact that $1 < r_2(\cRatio) < r_1(\cRatio)$ for all $\cRatio > 9$.
We find
\begin{align*}
((\cRatio - 5)\Dmin - t)r_2(\cRatio) &< ((\cRatio - 5)\Dmin - t) \\
((\cRatio - 5)\Dmin - t)r_2(\cRatio) - ((\cRatio - 3)\Dmin - t) &< ((\cRatio - 5)\Dmin - t) - ((\cRatio - 3)\Dmin - t) \\
((\cRatio - 5)\Dmin - t)r_2(\cRatio) - ((\cRatio - 3)\Dmin - t) &< -2\Dmin \\
((\cRatio - 5)\Dmin - t)r_2(\cRatio) - ((\cRatio - 3)\Dmin - t) &< 0 \\
\frac{r_2(\cRatio)(\cRatio - 1)\left( ((\cRatio - 5)\Dmin - t) \, r_2(\cRatio) - ((\cRatio - 3)\Dmin - t) \right)}{4(r_2(\cRatio)-1)(r_1(\cRatio)-r_2(\cRatio))} &< 0.
\end{align*}
\end{enumerate}

We now prove $a(\cRatio, t) \geq 0$
for all $\cRatio > 9$ and $t > 2\Dmin$ by contradiction.
Suppose that $a(\cRatio, t) < 0$.
Therefore,
\begin{align*}
\frac{d \delta}{d n} &= a(\cRatio, t)\log(r_1(\cRatio)) \, r_1^n(\cRatio) - b(\cRatio, t) \log(r_2(\cRatio))\, r_2^n(\cRatio) \\
&= b(\cRatio, t) \log(r_2(\cRatio))\, r_2^n(\cRatio)\left(\frac{a(\cRatio, t) \log(r_1(\cRatio)) }{b(\cRatio, t) \log(r_2(\cRatio))}\left(\frac{r_1(\cRatio)}{r_2(\cRatio)}\right)^n - 1\right) .
\end{align*}
Since $1 < r_2(\cRatio) < r_1(\cRatio)$, $b(\cRatio, t) < 0$ and $a(\cRatio, t) < 0$,
we have $\frac{a(\cRatio, t)}{b(\cRatio, t)} > 0$,
$\frac{\log(r_1(\cRatio))}{\log(r_2(\cRatio))} > 0$
and $\frac{r_1(\cRatio)}{r_2(\cRatio)} > 1$.
Therefore,
there exists a rank $n_0$ such that
$$\frac{a(\cRatio, t) \log(r_1(\cRatio)) }{b(\cRatio, t) \log(r_2(\cRatio))}\left(\frac{r_1(\cRatio)}{r_2(\cRatio)}\right)^n > 1$$
for all $n \geq n_0$.
This implies that $\frac{d \delta}{d n} < 0$ for all $n \geq n_0$.
Thus,
$\delta(n)$ is decreasing for all $n \geq n_0$.
This contradicts the fact that $\delta(n)$ is unbounded.
Therefore,
$a(\cRatio, t) \geq 0$
for all $\cRatio > 9$ and $t > 2\Dmin$.

Hence,
for any $\Dmin$ and $t > 2\Dmin$,
we are looking for the smallest value of $\cRatio$ such that
$$
\begin{cases}
\cRatio \geq \frac{t^2+3tx_1+(t+2x_1)\sqrt{t(t+2x_1)}}{tx_1} , & \cr
0 < \Dmin \leq x_1 \leq \frac{(\cRatio -1)\Dmin-t}{2} , & \cr
\end{cases}
$$
or equivalently,
$$
\begin{cases}
\cRatio \geq \frac{t^2+3tx_1+(t+2x_1)\sqrt{t(t+2x_1)}}{tx_1} , & \cr
\cRatio \geq \frac{2x_1 + t + \Dmin}{\Dmin} , & \cr
x_1 \geq \Dmin > 0 . & \cr
\end{cases}
$$
This optimization problem solves to
$\cRatio = 2\frac{\left(\frac{t}{2\Dmin}+2\right)\left(\frac{t}{2\Dmin}+\frac{1}{2}\right)}{\frac{t}{2\Dmin}}$ and $x_1 = \left( 2+\left(\frac{t}{2\Dmin}\right)^{-1} \right) \, \Dmin$.
%\begin{align}
%\label{sol cRatio}
%\cRatio &= %2\frac{\left(\frac{t}{2\Dmin}+2\right)\left(\frac{t}{2\Dmin}+\frac{1}{2}\right)}{\frac{t}%{2\Dmin}} , \\
%\label{sol x1}
%x_1 &= \left( 2+\left(\frac{t}{2\Dmin}\right)^{-1} \right) \, \Dmin . 
%\end{align}
%Indeed,
%$$\frac{t^2+3tx_1+(t+2x_1)\sqrt{t(t+2x_1)}}{tx_1} = \frac{2x_1 + t + \Dmin}{\Dmin}$$
%solves to~\eqref{sol x1},
%from which we get~\eqref{sol cRatio}.
Moreover,
since
$\left( 2+\left(\frac{t}{2\Dmin}\right)^{-1} \right) \, \Dmin > \Dmin$,
this strategies satisfies all the prescribed constraints.
\qed

%Reasoning about a general upper bound on $x_{n+1}$
%in order to find a lower bound on $\cRatio$
%(refer to~\eqref{eqn general equiv})
%is an idea that was introduced by Klein~\cite[Section 7.3.1]{klein1997algorithmische}.
%However,
%the analysis Klein makes of the upper bound on $x_{n+1}$
%does not work in our case.
%Indeed,
%the presence of a turn cost significantly changes the form of the algebraic expressions %involved in the analysis.

\section{A General Framework for Searching on a Line}
\label{section general framework searching line}

In this section, we consider an infinite family of Searching-on-a-Line problems.
Let $\cost_1(x) = \alpha_1 x+\beta_1$ be the cost of walking distance $x$ away from the origin
and $\cost_2(y) = \alpha_2 y+\beta_2$ be the cost of walking distance $y$ back to the origin.
For instance,
we have $\cost_1(x) = \cost_2(x) = x$
for the problem of searching on a line,
and we have $\cost_1(x) = x$ and $\cost_2(x) = x + t$
for the problem of searching on a line with turn cost.
For the rest of this section,
we suppose that $\alpha_1 \geq 0$, $\alpha_2 \geq 0$,
$\alpha_1 + \alpha_2 > 0$, $\beta_1 \geq 0$ and $\beta_2 \geq 0$.
\begin{theorem}
\label{thm framework search line}
If $\frac{3\beta_1+2\beta_2}{2(\alpha_1+\alpha_2)\Dmin} \leq 1$,
the strategy
$$x_i =\left(\left(\left(1-\frac{3\beta_1+2\beta_2}{2(\alpha_1+\alpha_2)\Dmin}\right)i+\left(1+\frac{\beta_1+\beta_2}{(\alpha_1+\alpha_2)\Dmin}\right)\right)2^i-\frac{\beta_1+\beta_2}{(\alpha_1+\alpha_2)\Dmin}\right)\Dmin$$
is optimal and has a competitive cost of $5\alpha_1+4\alpha_2$.

If $\frac{3\beta_1+2\beta_2}{2(\alpha_1+\alpha_2)\Dmin} \geq 1$,
the strategy
$x_i = 
\left(\left(1 + \frac{\beta_1+\beta_2}{(\alpha_1+\alpha_2)\Dmin}\right)\Phi^i -\frac{\beta_1+\beta_2}{(\alpha_1+\alpha_2)\Dmin}\right)\Dmin$,
where
$$\Phi = 1+\left(\frac{2\beta_1+\beta_2-(\alpha_1+\alpha_2)\Dmin+\sqrt{\left(2\beta_1+\beta_2\right)^2-\beta_2^2+\left(\beta_2+(\alpha_1+\alpha_2)\Dmin\right)^2}}{2(\alpha_1+\alpha_2)\Dmin}\right)^{-1},$$
%\begin{align*}
%x_i = 
%&\left(1 + \frac{\beta_1+\beta_2}{(\alpha_1+\alpha_2)\Dmin}\right)
%\left(1+\left(\frac{2\beta_1+\beta_2-(\alpha_1+\alpha_2)%\Dmin+\sqrt{\left(2\beta_1+\beta_2\right)^2-\beta_2^2+\left(\beta_2+(\alpha_1+\alpha_2)%\Dmin\right)^2}}{2(\alpha_1+\alpha_2)\Dmin}\right)^{-1}\right)^i \Dmin\\
%&-\frac{\beta_1+\beta_2}{\alpha_1+\alpha_2}
%\end{align*}
is optimal and has a competitive cost of
$\frac{(\alpha_1+\alpha_2)x_1+(\beta_1+\beta_2)+(\alpha_1\lambda+\beta_1)}{\lambda}$.
\end{theorem}

\proof
If $\beta_1 = \beta_2 = 0$,
the competitive cost of an optimal strategy is $5\alpha_1 + 4\alpha_2$.
That claim can be proven using Theorem~\ref{thm gal}.
Therefore,
the proof of the first statement of the theorem is identical
to the one of Lemma~\ref{lemma turn cost c = 9}
and the proof of the second statement is identical
to the one of Theorem~\ref{thm turn cost with m = 2}.
\qed

We can see where the ``$9$'' comes from in the original problem of searching on a line
by setting $\beta_1 = \beta_2 = 0$.
Notice that
Theorems~\ref{thm gal} and~\ref{thm framework search line}
to solutions for two different infinite families of search problems.
%Moreover,
%by looking at the general expressions for $x_i$ and for the competitive cost of the optimal %strategy,
%we see that there is no symmetry bewteen moving away from the origin and getting closer to the %origin. This can be explained by the fact that
%the number of times we move away from the origin is exactly once more
%than the number of times we move towards the origin.

\section{Searching on $m$ Rays with Turn Cost}
\label{section searching m rays turn cost}

In this section,
we consider the problem of searching on $m$ rays extending from the origin where the cost is measured as the total distance travelled plus $t \geq 0$ times the number of turns
We suppose that a lower bound $\Dmin \leq D$ is given to the searcher.
We have the following result.
\begin{theorem}
\label{thm turn cost with general m}
If
$\frac{t}{2\Dmin} \leq \frac{1}{\left(\frac{m}{m-1}\right)^{m-1}-1}$,
the strategy
$$x_i = \left(\left(\frac{1}{m-1}\left(1 - \left(\left(\frac{m}{m-1}\right)^{m-1}-1\right)\frac{t}{2\Dmin} \right) i +\left(1+\frac{t}{2\Dmin}\right)\right)\left(\frac{m}{m-1}\right)^i - \frac{t}{2\Dmin}\right)\Dmin $$
is optimal and has a competitive cost of $1+2\frac{m^m}{(m-1)^{m-1}}$.

If
$\frac{t}{2\Dmin} \geq \frac{1}{\left(\frac{m}{m-1}\right)^{m-1}-1}$,
the strategy
$$x_i = \left(\left(1+\frac{t}{2\Dmin}\right)\left(1+\left(\frac{t}{2\Dmin}\right)^{-1}\right)^{\frac{i}{m-1}} - \frac{t}{2\Dmin}\right)\Dmin $$
has a competitive cost of
$\frac{\left(1+\left(\frac{t}{2\Dmin}\right)^{-1}\right)^{\frac{-1}{m-1}} - \left(3 + 2\left(\frac{t}{2\Dmin}\right)^{-1}\right)}{\left(1+\left(\frac{t}{2\Dmin}\right)^{-1}\right)^{\frac{-1}{m-1}} - 1}$.
\end{theorem}

\proof
If
$\frac{t}{2\Dmin} \leq \frac{1}{\left(\frac{m}{m-1}\right)^{m-1}-1}$,
$$\frac{\sum_{i=1}^{m-1}(2x_i + t) + \Dmin}{\Dmin} = 1+2\frac{m^m}{(m-1)^{m-1}}$$
and
$$\frac{\sum_{i=1}^{n+(m-1)}(2x_i + t) + x_n}{x_n} = 1+2\frac{m^m}{(m-1)^{m-1}}$$
for all $n \geq 1$.
Therefore,
it is optimal.

If $\frac{t}{2\Dmin} \geq \frac{1}{\left(\frac{m}{m-1}\right)^{m-1}-1}$,
$$\frac{\sum_{i=1}^{m-1}(2x_i + t) + \Dmin}{\Dmin} = \frac{\left(1+\left(\frac{t}{2\Dmin}\right)^{-1}\right)^{\frac{-1}{m-1}} - \left(3 + 2\left(\frac{t}{2\Dmin}\right)^{-1}\right)}{\left(1+\left(\frac{t}{2\Dmin}\right)^{-1}\right)^{\frac{-1}{m-1}} - 1}$$
and
$$\frac{\sum_{i=1}^{n+(m-1)}(2x_i + t) + x_n}{x_n} = \frac{\left(1+\left(\frac{t}{2\Dmin}\right)^{-1}\right)^{\frac{-1}{m-1}} - \left(3 + 2\left(\frac{t}{2\Dmin}\right)^{-1}\right)}{\left(1+\left(\frac{t}{2\Dmin}\right)^{-1}\right)^{\frac{-1}{m-1}} - 1}$$
for all $n \geq 1$.
\qed

\subsection{Conjecture and Open Problems}
We conjecture that the strategy of Theorem~\ref{thm turn cost with general m} is optimal for all $t \geq 0$. Our
belief is based on the following.
Suppose we replace the inequalities
in~\eqref{basic ineq x1} and~\eqref{basic ineq xi}
by equalities.
The system of equations we obtain defines a unique strategy $x_i^*$,
which depends on $\cRatio$.
Suppose we find the smallest value of $\cRatio$ such that $x_i^*$
is fully monotonic and unbounded.
There is no guaranty that $x_i^*$ is optimal
since we restricted the set of possible strategies
by replacing the inequalities by equalities.
However,
it turns out that $x_i^*$ is the strategy of Theorem~\ref{thm turn cost with m = 2}.
And since the strategy of Theorem~\ref{thm turn cost with m = 2} is optimal,
then $x_i^*$ is optimal.
The same is true for the general framework of Theorem~\ref{thm framework search line}.

We are unable to prove a version of Theorem~\ref{thm turn cost with m = 2} for $m>2$.
However,
if we replace the inequalities by equalities and we optimize for $\cRatio$,
we find the strategy of Theorem~\ref{thm turn cost with general m}.
Moreover,
the strategy of Theorems~\ref{thm turn cost with m = 2}
and~\ref{thm framework search line} are of the form $x_i=(a\,i+b)c^i+d$,
for some constants $a$, $b$, $c$ and $d$.
When $m > 2$,
if we search for a strategy of the form $x_i=(a\,i+b)c^i+d$ that minimizes $\cRatio$,
we get the strategy of Theorem~\ref{thm turn cost with general m}.
For all these reasons,
we conjecture the strategy of Theorem~\ref{thm turn cost with general m}
is optimal.

There are two main open problems remaining. Prove our conjecture about the problem of searching on $m$ rays with turn cost. Prove a version of Theorem~\ref{thm framework search line} for $m > 2$.

\newpage
\bibliographystyle{plain}

\appendix
\section{Monotonic, Fully Monotonic and Periodic Strategies}
\label{section monotonic periodic}

In this section,
we provide a proof for Theorem~\ref{thm general periodic monotonic m = 2}.
Refer to Section~\ref{section previous work} for notations.
We know from previous work
(see~\cite{DBLP:journals/iandc/Baeza-YatesCR93,gal1980} for instance)
that there is an optimal strategy that is periodic and monotonic.
However,
we want to determine what properties $\cost_1(\cdot)$ and $\cost_2(\cdot)$ must
satisfy so that there exists an optimal search strategy that is periodic and fully monotonic.
To establish as weak constraints as possible on $\cost_1(\cdot)$ and $\cost_2(\cdot)$,
we look closely at all details of the proof.
%At the end of this section,
%we also present a review of different proofs that were published.

For a searcher to find the target at step $j$,
$\mathcal{S}$ and the target must satisfy the following properties.
\begin{itemize}
\item The target is on $r_j$.

\item The distance $D$ between the target and the origin is such that $x_{j'} < D \leq x_j$ for all $j' < j$ such that $r_{j'} = r_j$.
\end{itemize}
When,
$x_{j'} < x_j$ for all $j' < j$ such that $r_{j'} = r_j$,
we say that $j$ is \emph{feasible for} $\mathcal{S}$.
We also define $\prev(\mathcal{S},j) < j$ to be the index such that $r_{\prev(\mathcal{S},j)} = r_j$ and $x_{\prev(\mathcal{S},j)}$ is the largest distance that was
travelled on $r_j$ during the first $j-1$ steps.
If this is the first time that $r_j$ is visited,
then let $\prev(\mathcal{S},j) = 0$ and $x_{\prev(\mathcal{S},j)} = x_0 = \Dmin$.
Notice that $x_0$ is defined to simplify the presentation of the proofs in this section.
The first step of $\mathcal{S}$ is $\mathcal{S}(1) = (x_1,r_1)$.
Consider a strategy $\mathcal{S}$ such that $\Dmin = 1$, $\mathcal{S}(1) = (6,left)$, $\mathcal{S}(2) = (3,right)$, $\mathcal{S}(3) = (2,left)$, $\mathcal{S}(4) = (4,right)$, $\mathcal{S}(5) = (5,right)$ and $\mathcal{S}(6) = (3,left)$.
We have that $3$ and $6$ are not feasible for $\mathcal{S}$ since it is impossible for the target to be discovered at Steps $3$ or $6$.
On the other hand,
$1$, $2$, $4$ and $5$ are feasible for $\mathcal{S}$.
We have $\prev(\mathcal{S},1)= 0$,
$\prev(\mathcal{S},2)= 0$,
$\prev(\mathcal{S},3)= 1$,
$\prev(\mathcal{S},4)= 2$,
$\prev(\mathcal{S},5)= 4$
and $\prev(\mathcal{S},6)= 1$.

To characterize how good a strategy $\mathcal{S}$ is,
we compute the competitive distance the searcher needs to walk before finding the target in the worst case,
by following $\mathcal{S}$.
In the worst case,
the cost of finding the target at step $j$
(for a feasible $j$)
is
\begin{align}
\label{definition CRj}
CR_j(\mathcal{S}) = \sup_{x_{\prev(\mathcal{S},j)} < D \leq x_j} \frac{\sum_{i=1}^{j-1} 2x_i + D}{D} = \frac{\sum_{i=1}^{j-1} 2x_i + x_{\prev(\mathcal{S},j)}}{x_{\prev(\mathcal{S},j)}}.
\end{align}
And the competitive distance the searcher needs to walk before finding the target in the worst case is
$$CR(\mathcal{S}) = \sup_{j \geq 1}CR_{j}(\mathcal{S}) ,$$
where the supremum is taken over all feasible steps $j$.

Consider for instance
the \emph{Power of Two} strategy
$\mathcal{S}(i) = \left(2^i\Dmin,r_i\right)$,
where $r_{2 k - 1} = \textrm{left}$
and $r_{2 k} = \textrm{right}$ for all $k\geq 1$.
Then all steps are feasible and
the competitive cost is
$$CR(\mathcal{S}) = \sup_{j\geq 1} CR_j(\mathcal{S}) = \sup_{j\geq 1} \left(9 - 2^{4-j}\right) = 9.$$

\begin{definition}[Monotonic and Fully Monotonic Strategies]
Let $\mathcal{S}(i) = (x_i,r_i)$ be a search strategy.
We say that $\mathcal{S}$ is \emph{monotonic}
if the sequences $\left(x_i\right)_{Left}$
and $\left(x_i\right)_{Right}$
are strictly increasing.
We say that $\mathcal{S}$ is \emph{fully monotonic}
if it is monotonic and if the sequence $\left(x_i\right)_{i\geq 1}$ is monotonic and non-decreasing.
\end{definition}

We prove that,
without loss of generality,
we can suppose that any search strategy is monotonic.
\begin{lemma}
\label{lemma monotonic}
Let $\mathcal{S} = (x_i,r_i)$ be a search strategy with competitive ratio $\cRatio$.
There exists a monotonic search strategy $\mathcal{S}^* = (x_i^*,r_i^*)$
with competitive ratio at most $\cRatio$.
\end{lemma}

\proof
If $\mathcal{S}$ is monotonic,
there is nothing to prove.
Suppose that $\mathcal{S}$ is not monotonic.
Therefore,
without loss of generality,
the sequence $\left(x_i\right)_{i\in Left}$ is not strictly increasing.
Hence,
we can consider the two smallest integers $k,k'\in Left$ such that $k<k'$ and $x_k \geq x_{k'}$.
Let
$$\mathcal{S}'(i) = \begin{cases} S(i) & i < k', \cr S(i+1) & i \geq k'. \end{cases}$$
We show that $CR(\mathcal{S}') \leq \cRatio$.
Notice that $\mathcal{S}'$ satisfies $\sup_{i\in Left}(x'_i) = \sup_{i\in Right}(x'_i) = \infty$.

Let $j\geq 1$ be an integer and suppose that
$j$ is feasible for $\mathcal{S}'$.
We consider two cases:
either (1) $j < k'$
or (2) $j \geq k'$.
\begin{enumerate}
\item 
If $j < k'$,
then,
since $j$ is feasible for $\mathcal{S}'$,
we have $x'_j > x'_{\prev(\mathcal{S}',j)}$.
Moreover,
since $j < k'$,
we have $x_j = x'_j$ and $x_{\prev(\mathcal{S},j)} = x'_{\prev(\mathcal{S}',j)}$,
from which we get $x_j > x_{\prev(\mathcal{S},j)}$.
Therefore,
$j$ is feasible for $\mathcal{S}$.
Then,
the competitive cost 
of finding the target at step $j$ with $\mathcal{S}'$ is
\begin{align*}
CR_j(\mathcal{S}') &= \frac{\sum_{i=1}^{j-1} 2x'_i + x'_{\prev(\mathcal{S}',j)}}{x'_{\prev(\mathcal{S}',j)}}\\
&= \frac{\sum_{i=1}^{j-1} 2x_i + x_{\prev(\mathcal{S},j)}}{x_{\prev(\mathcal{S},j)}} & \textrm{since $j < k'$,}\\
&= CR_j(\mathcal{S}) .
\end{align*}

\item 
If $j \geq k'$,
then,
since $j$ is feasible for $\mathcal{S}'$,
we have $x'_j > x'_{\prev(\mathcal{S}',j)}$.
Moreover,
since $j \geq k'$,
we have $x_{j+1} = x'_j$ and $x_{\prev(\mathcal{S},j+1)} = x'_{\prev(\mathcal{S}',j)}$,
from which we get $x_{j+1} > x_{\prev(\mathcal{S},j+1)}$.
Therefore,
$j+1$ is feasible for $\mathcal{S}$.
Then,
the competitive cost 
of finding the target at step $j$ with $\mathcal{S}'$ is
\begin{align*}
CR_j(\mathcal{S}') &= \frac{\sum_{i=1}^{j-1} 2x'_i + x'_{\prev(\mathcal{S}',j)}}{x'_{\prev(\mathcal{S}',j)}}\\
&= \frac{\sum_{i=1}^j 2x_i - 2x_{k'} + x'_{\prev(\mathcal{S}',j)}}{x'_{\prev(\mathcal{S}',j)}} & \textrm{since $j \geq k'$,}\\
&= \frac{\sum_{i=1}^j 2x_i - 2x_{k'} + x_{\prev(\mathcal{S},j+1)}}{x_{\prev(\mathcal{S},j+1)}} \\
&\leq \frac{\sum_{i=1}^j 2x_i + x_{\prev(\mathcal{S},j+1)}}{x_{\prev(\mathcal{S},j+1)}}\\
&= CR_{j+1}(\mathcal{S}) .
\end{align*}
\end{enumerate}

Consequently,
in both cases,
$$CR(\mathcal{S}') = \sup_{j\geq 1}CR_j(\mathcal{S}') \leq \sup_{j\geq 1}CR_j(\mathcal{S}) = CR(\mathcal{S}) = \cRatio ,$$
where the suprema are taken over all feasible steps $j$.
We repeat the same transformation on each non-monotonic part of $\mathcal{S}$.
This leads to a monotonic search strategy $\mathcal{S}^*$.
\qed

We use Lemma~\ref{lemma monotonic} to prove that,
without loss of generality,
we can suppose that any search strategy
is \emph{periodic} and fully monotonic.
\begin{definition}[Periodic Strategy]
Let $\mathcal{S}(i) = (x_i,r_i)$ be a search strategy.
We say that $\mathcal{S}$ is \emph{periodic}
if $r_1\neq r_2$ and $r_i = r_{i+2}$ for all $i\geq 1$.
\end{definition}

\begin{lemma}
\label{lemma periodic fully monotonic}
Let $\mathcal{S} = (x_i,r_i)$ be a monotonic search strategy with competitive ratio $\cRatio$.
There exists a search strategy $\mathcal{S}' = (x'_i,r'_i)$,
with competitive ratio at most $\cRatio$,
that is periodic and fully monotonic.
\end{lemma}

\proof
Let $(x'_i)_{i\geq 1}$ be the sequence obtained from the sequence $(x_i)_{\geq 1}$ by sorting it in non-decreasing order.
Therefore,
\begin{align}
\label{ineq sorted seq}
\sum_{i=1}^{j-1} 2x'_i \leq \sum_{i=1}^{j-1} 2x_i
\end{align}
for all $j \geq 1$.
Let $\mathcal{S}'(i) = (x'_i,r'_i)$,
where $r'_{2 k - 1} = \textrm{left}$
and $r'_{2 k} = \textrm{right}$ for all $k\geq 1$.
We show that $CR(\mathcal{S}') \leq \cRatio$.
Notice that $\mathcal{S}'$ satisfies $\sup_{i\in Left}(x'_i) = \sup_{i\in Right}(x'_i) = \infty$.

Let $j\geq 1$ be an integer and suppose that
$j$ is feasible for $\mathcal{S}'$.
We consider two cases:
either (1) there exists a $t \geq j-1$ such that $x_t \leq x'_{j-2}$
or (2) for all $t \geq j-1$, $x_t > x'_{j-2}$.
\begin{enumerate}
\item Let $t' > t \geq j - 1$ be the smallest index such that $r_{t'} = r_t$.
Since $\mathcal{S}$ is monotonic,
we have $x_{t'} > x_t$.
Consider the scenario where we place a target on $r_t$ at a distance $D$ such that $x_t < D \leq x_{t'}$.
In such a situation,
by following $\mathcal{S}$,
the searcher finds the target at step $t'$.
Moreover,
from~\eqref{definition CRj},
we have
\begin{align}
\label{eq case 1 CR t'}
CR_{t'}(\mathcal{S}) = \frac{\sum_{i=1}^{t'-1} 2x_i + x_t}{x_t} .
\end{align}
We get
\begin{align*}
CR_j(\mathcal{S}') &= \frac{\sum_{i=1}^{j-1} 2x'_i + x'_{\prev(\mathcal{S}',j)}}{x'_{\prev(\mathcal{S}',j)}} \\
&= \frac{\sum_{i=1}^{j-1} 2x'_i + x'_{j-2}}{x'_{j-2}} &\textrm{since $\mathcal{S}'$ is periodic,}\\
&\leq \frac{\sum_{i=1}^{j-1} 2x_i + x'_{j-2}}{x'_{j-2}} &\textrm{by~\eqref{ineq sorted seq},}\\
&\leq \frac{\sum_{i=1}^{j-1} 2x_i + x_t}{x_t} &\textrm{since $x_t \leq x'_{j-2}$,}\\
&\leq \frac{\sum_{i=1}^{t'-1} 2x_i + x_t}{x_t} &\textrm{since $t'-1 \geq t\geq j-1$,}\\
&= CR_{t'}(\mathcal{S}) &\textrm{by~\eqref{eq case 1 CR t'}.}
\end{align*}

\item In this case,
\begin{align}
\label{eq sets}
\{x_1,x_2,...,x_{j-2}\} = \{x'_1,x'_2,...,x'_{j-2}\}.
\end{align}
In other words,
the sequence $(x_i)_{1 \leq i \leq j-2}$ is a permutation
of the sequence $(x'_i)_{1 \leq i \leq j-2}$.
We subdivide this case into two subcases:
either (a) $r_k = r_1$ for all $1 \leq k \leq j-2$
or (b) not.
\begin{enumerate}
\item Together with~\eqref{eq sets},
since $\mathcal{S}$ is monotonic,
we have
$(x_i)_{1 \leq i \leq j-2} = (x'_i)_{1 \leq i \leq j-2}$.
In particular,
we have
\begin{align}
\label{eq x j-2}
x_{j-2} = x'_{j-2}.
\end{align}
Let $t' > j-2$ be the smallest index such that $r_{t'} \neq r_{j-2}$.
Either we have i. $t' = j-1$ or ii. $t' > j-1$.
\begin{enumerate}
\item If $t' = j-1$,
let $t'' > j-2$ be the smallest index such that $r_{t''} = r_{j-2}$.
Since $r_{t'} \neq r_{j-2}$,
we have $t'' > t' = j-1$.
Also,
since $\mathcal{S}$ is monotonic,
we have $x_{t''} > x_{j-2}$.
Consider the scenario where we place a target on $r_{j-2}$ at a distance $D$ such that $x_{j-2} < D \leq x_{t''}$.
In such a situation,
by following $\mathcal{S}$,
the searcher finds the target at step $t''$.
Moreover,
from~\eqref{definition CRj},
we have
$$CR_{t''}(\mathcal{S}) = \frac{\sum_{i=1}^{t''-1} 2x_i + x_{j-2}}{x_{j-2}} .$$
We get
\begin{align*}
CR_j(\mathcal{S}') &= \frac{\sum_{i=1}^{j-1} 2x'_i + x'_{\prev(\mathcal{S}',j)}}{x'_{\prev(\mathcal{S}',j)}} \\
&= \frac{\sum_{i=1}^{j-1} 2x'_i + x'_{j-2}}{x'_{j-2}} &\textrm{since $\mathcal{S}'$ is periodic,}\\
&\leq \frac{\sum_{i=1}^{j-1} 2x_i + x'_{j-2}}{x'_{j-2}} &\textrm{by~\eqref{ineq sorted seq},}\\
&= \frac{\sum_{i=1}^{j-1} 2x_i + x_{j-2}}{x_{j-2}} &\textrm{by~\eqref{eq x j-2},}\\
&\leq \frac{\sum_{i=1}^{t''-1} 2x_i + x_{j-2}}{x_{j-2}} &\textrm{since $t''-1 \geq t' = j-1$,}\\
&= CR_{t''}(\mathcal{S}).
\end{align*}

\item For the case where $t' > j-1$,
consider the scenario where we place a target on $r_{t'}$ at distance $D$ such that $\Dmin < D \leq x_{t'}$.
In such a situation,
by following $\mathcal{S}$,
the searcher finds the target at step $t'$.
Moreover,
from~\eqref{definition CRj},
we have
$$CR_{t'}(\mathcal{S}) = \frac{\sum_{i=1}^{t'-1} 2x_i + \Dmin}{\Dmin} .$$
We get
\begin{align*}
CR_j(\mathcal{S}') &= \frac{\sum_{i=1}^{j-1} 2x'_i + x'_{\prev(\mathcal{S}',j)}}{x'_{\prev(\mathcal{S}',j)}} \\
&= \frac{\sum_{i=1}^{j-1} 2x'_i + x'_{j-2}}{x'_{j-2}} &\textrm{since $\mathcal{S}'$ is periodic,}\\
&\leq \frac{\sum_{i=1}^{j-1} 2x_i + x'_{j-2}}{x'_{j-2}} &\textrm{by~\eqref{ineq sorted seq},}\\
&\leq \frac{\sum_{i=1}^{j-1} 2x_i + \Dmin}{\Dmin} &\textrm{since $\Dmin \leq x'_{j-2}$,}\\
&\leq \frac{\sum_{i=1}^{t'-1} 2x_i + \Dmin}{\Dmin} &\textrm{since $t'-1 \geq j-1$,}\\
&= CR_{t'}(\mathcal{S}).
\end{align*}
\end{enumerate}

\item In this case,
there exist a largest index $j_{\textrm{left}} \leq j-2$ such that $r_{j_{\textrm{left}}} = \textrm{left}$
and a largest index $j_{\textrm{right}} \leq j-2$ such that $r_{j_{\textrm{right}}} = \textrm{right}$.
From~\eqref{eq sets},
we get $x_{j_{\textrm{left}}} \leq x'_{j-2}$
and $x_{j_{\textrm{right}}} \leq x'_{j-2}$.
Let $j_{\textrm{left}}' > j-2$ be the smallest index such that $r_{j_{\textrm{left}}'} = \textrm{left}$ and
let $j_{\textrm{right}}' > j-2$ be the smallest index such that $r_{j_{\textrm{right}}'} = \textrm{right}$.
Since $j_{\textrm{left}}' \neq j_{\textrm{right}}'$,
we have $j_{\textrm{left}}' > j - 1$
or $j_{\textrm{right}}' > j - 1$.
Without loss of generality,
suppose we have $j_{\textrm{left}}' > j - 1$.
Since $\mathcal{S}$ is monotonic,
we have $x_{j_{\textrm{left}}'} > x_{j_{\textrm{left}}}$.

Consider the scenario where we place a target on the left ray at distance $D$ such that $x_{j_{\textrm{left}}} < D \leq x_{j_{\textrm{left}}'}$.
In such a situation,
by following $\mathcal{S}$,
the searcher finds the target at step $j_{\textrm{left}}'$.
Moreover,
from~\eqref{definition CRj},
we have
$$CR_{j_{\textrm{left}}'}(\mathcal{S}) = \frac{\sum_{i=1}^{j_{\textrm{left}}'-1} 2x_i + x_{j_{\textrm{left}}}}{x_{j_{\textrm{left}}}} .$$
We get
\begin{align*}
CR_j(\mathcal{S}') &= \frac{\sum_{i=1}^{j-1} 2x'_i + x'_{\prev(\mathcal{S}',j)}}{x'_{\prev(\mathcal{S}',j)}} \\
&= \frac{\sum_{i=1}^{j-1} 2x'_i + x'_{j-2}}{x'_{j-2}} &\textrm{since $\mathcal{S}'$ is periodic,}\\
&\leq \frac{\sum_{i=1}^{j-1} 2x_i + x'_{j-2}}{x'_{j-2}} &\textrm{by~\eqref{ineq sorted seq},}\\
&\leq \frac{\sum_{i=1}^{j-1} 2x_i + x_{j_{\textrm{left}}}}{x_{j_{\textrm{left}}}} &\textrm{since $x_{j_{\textrm{left}}} \leq x_{j-2}'$,}\\
&\leq \frac{\sum_{i=1}^{j_{\textrm{left}}'-1} 2x_i + x_{j_{\textrm{left}}}}{x_{j_{\textrm{left}}}} &\textrm{since $j_{\textrm{left}}'-1 \geq j-1$,}\\
&= CR_{j_{\textrm{left}}'}(\mathcal{S}).
\end{align*}
\end{enumerate}
\end{enumerate}
Consequently,
in all cases,
$$CR(\mathcal{S}') = \sup_{j\geq 1}CR_j(\mathcal{S}') \leq \sup_{j\geq 1}CR_j(\mathcal{S}) = CR(\mathcal{S}) = \cRatio ,$$
where the suprema are taken over all feasible steps $j$.
\qed

From Lemmas~\ref{lemma monotonic} and~\ref{lemma periodic fully monotonic},
we deduce the following corollary.
\begin{corollary}
\label{corollary at last}
There exists an optimal search strategy that is periodic and fully monotonic.
\end{corollary}

Theorem~\ref{thm general periodic monotonic m = 2} is now a direct consequence of Corollary~\ref{corollary at last}.

\proof (Theorem~\ref{thm general periodic monotonic m = 2})
If we replace the definition of $CR_j(\cdot)$
(refer to~\eqref{definition CRj}) by
$$CR_j^ *(\mathcal{S}) = \frac{\sum_{i=1}^{j-1} (\cost_1(x_i)+\cost_2(x_i)) + \cost_1\left(x_{\prev(\mathcal{S},j)}\right)}{x_{\prev(\mathcal{S},j)}}$$
in the proof of Lemmas~\ref{lemma monotonic} and~\ref{lemma periodic fully monotonic},
the exact same proof stands,
and a fortiori the proof of Corollary~\ref{corollary at last} stands.
\qed

%\subsection{A review of Alternative Proofs}
%
%
%IN ORDER TO PROVE THAT 9 IS OPTIMAL, the following authors first prove...
%
%
%In~\cite{DBLP:journals/iandc/Baeza-YatesCR93},
%Baeza-Yates et al. suppose periodicity without explicit mention.
%They explain why there is an optimal solution that is monotonic
%(no formal proof, but no real need to).
%Don't suppose or prove fully monotonic.
%
%In~\cite{DBLP:journals/tcs/Lopez-OrtizS01},
%L\'opez-Ortiz and Schuierer work on
%in a framework where the searcher is given an upper bound $\Dmax$ on $D$
%such that $D \leq \Dmax$.
%
%In~\cite{langetepeThesis},
%Langetepe explains why there is an optimal solution that is periodic (no formal proof,
%but no real need to).
%He also explains why there is an optimal solution that is monotonic (no formal proof,
%but no real need to).
%Then he proves that there is an optimal solution that is fully monotonic,
%but for the setting where there is an upper bound on $D$ that is known to the searcher.
%
%In~\cite{gal1980},
%formal proof for periodic and monotonic.
%
%In~\cite{yin1994},
%Yin works in a framework where multiple robots are searching for a target on $m$ rays.
%

\section{Proof of Lemma~\ref{lemma > 9}}
\label{section proof lemma}

In this section,
we provide a proof for Lemma~\ref{lemma > 9}.

\proof (Lemma~\ref{lemma > 9})
We prove Lemma~\ref{lemma > 9} by contradiction.
Nonetheless,
the proof is similar to that of Theorem~\ref{thm turn cost with m = 2}.
Suppose that $\frac{t}{2\Dmin} > 1$ and $\cRatio = 9$
and let $x_i$ be an optimal strategy.
We know that $x_i$ satisfies
$\frac{(2x_1 + t) + \Dmin}{\Dmin} \leq 9$
and
\begin{align}
\label{basic ineq xi lemma}
\frac{\sum_{i=1}^{n+1} (2x_i + t) + x_n}{x_n} \leq 9
\end{align}
for all $n\geq 1$.
Therefore,
\begin{align}
\label{ineq constraint x1 lemma}
0 < \Dmin \leq x_1 \leq \frac{8\Dmin - t}{2}
\end{align}
and for all $n \geq 1$,
\begin{align}
\label{eqn general equiv lemma}
x_{n+1} \leq 3 x_n - t - \sum\limits_{i=1}^{n-1} \left(x_i + \frac{t}{2}\right) .
\end{align}
We can prove by induction that
for all $0 \leq m \leq n-1$,
\begin{align}
\label{eqn f n+1 m lemma}
x_{n+1} \leq \tau_m x_{n-m} - \mu_m - \nu_m\sum\limits_{i=1}^{n-1 - m} (x_i + C),
\end{align}
where
\begin{align*}
\tau_0 &= 3 , & \mu_0 &= t , & \nu_0 &= 1 , \\
\tau_{m+1} &= 3\,\tau_m - \nu_m , & \mu_{m+1} &= t\,\tau_m + \mu_m + \frac{t}{2}\,\nu_m , & \nu_{m+1} &= \tau_m + \nu_m .
\end{align*}
From the theory of characteristic equations,
we find
$$\tau_n = (n+3)2^n , \qquad \mu_n = \left((3n+1)2^n+1\right)\frac{t}{2} , \qquad \nu_n = (n+1)2^n .$$

From~\eqref{eqn f n+1 m lemma} with $m:=n-1$,
we have
\begin{align}
\label{ineq Klein lemma}
x_{n+1} \leq \tau_{n-1} x_1 - \mu_{n-1}
\end{align}
for all $n \geq 1$.
By Theorem~\ref{thm general periodic monotonic m = 2},
$x_{n+1}$ is increasing with respect to $n$.
Also,
$x_{n+1}$ is unbounded.
Therefore,
$\delta(n) = \tau_{n-1} x_1 - \mu_{n-1}$
must be unbounded and increasing for all $n\geq 1$.

The function $\delta(n)$ is increasing for all $n\geq 1$
if and only if 
\begin{align}
\label{ineq proof lemma c > 9}
\delta(n+1)-\delta(n) = \frac{2(n+4)x_1 -(3n+4)t}{4}\,2^n > 0
\end{align}
for all $n \geq 1$.
Therefore,
we must have
$t < \frac{2(n+4)}{3n+4}\,x_1$
for all $n \geq 1$.
By letting $n\rightarrow\infty$,
we find that $t$ must satisfy
$t \leq \frac{2}{3}\,x_1$.
Since $x_1 \leq \frac{8\Dmin - t}{2}$ from~\eqref{basic ineq xi lemma},
we find
$t \leq \frac{2}{3}\,x_1 \leq \frac{8\Dmin - t}{3}$,
from which $t \leq 2\Dmin$,
which is a contradiction.
Consequently,
$\cRatio > 9$.
\qed

\section{Figures for Sections~\ref{subsubsection tradeoff} and~\ref{section different work}}
\label{section figures}

\begin{figure}[h]
\centering
\includegraphics[scale=1]{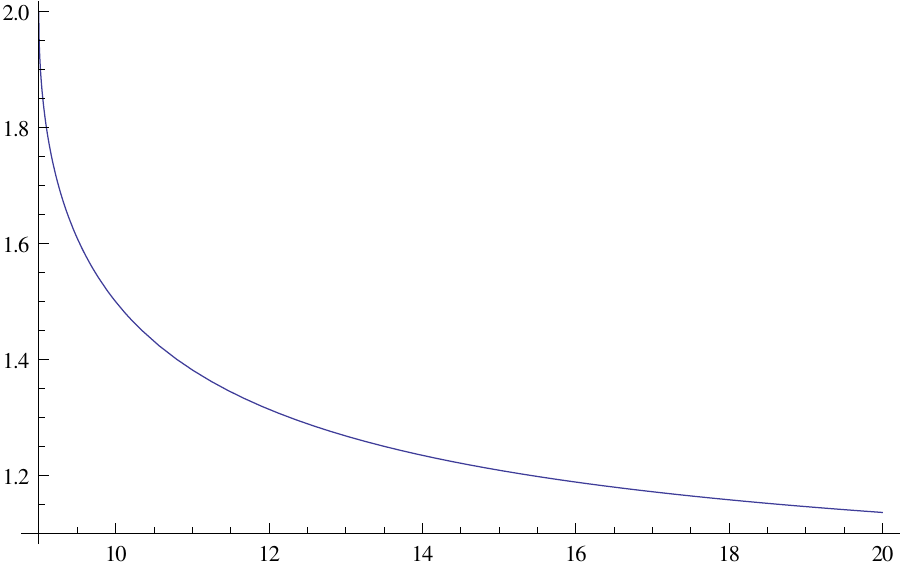}
\caption{Tradeoff between $\cRatio$ and $\frac{1}{t}\phi(\cRatio)$
(see Section~\ref{subsubsection tradeoff}).\label{figure tradeoff}}
\end{figure}

\begin{figure}[h]
\centering
\includegraphics[scale=1]{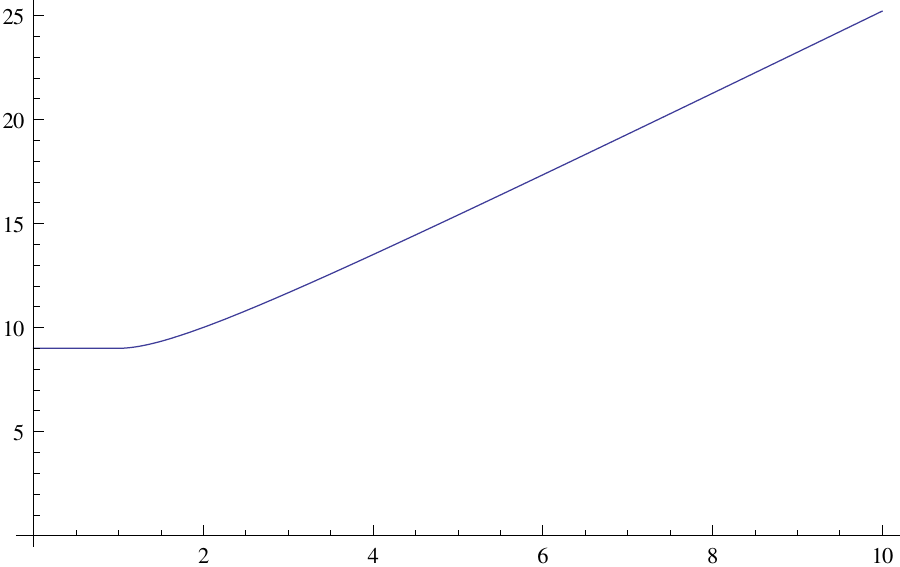}
\caption{Optimal cost with respect to $\frac{t}{2\Dmin}$
(see Section~\ref{section different work}).\label{figure optimal with turn cost}}
\end{figure}

\end{document}